\documentclass[prb,aps,twocolumn,showpacs,preprintnumbers,amsmath,amssymb]{revtex4-1}

\usepackage{graphicx}
\usepackage{dcolumn}
\usepackage{color}
\usepackage{bm}
\usepackage{subfigure}

\begin{document}

\title{Thermodynamic properties of liquid mercury to 520~K and 7~GPa from acoustic velocity measurements}

\author{S.~Ayrinhac$^{1}$, M.~Gauthier$^{1}$, L.E.~Bove$^{1,2}$, M.~Morand$^{1}$, G.~Le~Marchand$^{1}$, F.~Bergame$^{1}$ and F.~Decremps$^{1}$}
\affiliation{$^{1}$Institut de Min\'{e}ralogie et Physique des Milieux Condens\'{e}s, Universit\'{e} Pierre et Marie Curie, 75252 Paris, France\\
$^{2}$Ecole Polytech. Fed. Lausanne, Inst. Condensed Matter Phys., EPSL, CH-1015 Lausanne, Switzerland}

\date{\today}

\begin{abstract}
Ultrafast acoustics measurements on liquid mercury have been performed at high pressure and temperature in diamond anvils cell using picosecond acoustic interferometry. 
We extract the density of mercury from adiabatic sound velocities using a numerical iterative procedure.  
The pressure and temperature dependence of the thermal expansion ($\alpha_{P}$), the isothermal compressibilty ($\beta_{T}$), the isothermal bulk modulus ($B_{T}$) and its pressure derivative ($B_{T}^{'}$) are derived up to 7~GPa and 520~K.
In the high pressure regime, the sound velocity values, at a given density, are shown to be only slightly dependent on the specific temperature and pressure conditions.
The density dependence of sound velocity at low density is consistent with that observed with our data at high density in the metallic liquid state.
\end{abstract}


\maketitle

For most of its thermodynamic properties liquid mercury can be described, as a simple liquid~\cite{Ingebrigtsen2012}, though it is a very unusual element compared to other close-shell elements. As an example, it is the only metal liquid at ambient conditions, due to relativistic effects on the core electrons~\cite{Norrby1991, Calvo2013}, and it exhibits anomalous electronic properties~\cite{Jank1990} compared to others transitional metals.

At low densities it undergoes a gradual metal (M) non nonmetal (NM) transition due to the lack of overlapping between the 6s and 6p bands~\cite{Kohno1999}. This transition occurring at a density around 9~g/cm$^{3}$ has been largely investigated both theoretically and experimentally~\cite{Edwards1995}. In correspondence to this transition the density dependence of the sound velocity in liquid mercury shows an abrupt change~\cite{Suzuki1980, Munejiri1998} which has been related to the modification of the interatomic interaction~\cite{Inui2003} when the non metallic state is attained. 

While the low-density regime has been widely studied~\cite{Suzuki1980, Munejiri1998}, the properties of liquid mercury at high densities are not well known.
At high densities the repulsive part of the pair potential function mainly determines the sound propagation velocity ~\cite{Bomont2006}, thus the study of liquid mercury under compression can provide interesting insights on the short range part of the interatomic interaction~\cite{Bove2002}. 
Furthermore, one of the most fundamental properties, i.e. the pressure-volume-temperature relation, which characterizes the thermodynamic equilibrium state, can be derived from the measurements of sound velocities as a function of pressure. Knowledge of this relation in the high density regime is here relevant to constrain the results of theoretical simulations on mercury and thus to improve the model of the effective pair potential function~\cite{Bomont2006, Bove2002, Munejiri1998}.

Unfortunately the measurement of the density at high pressures in equilibrium liquid state is technically demanding~\cite{Funtikov2009}, which explains why data in liquids at high pressure are still scarce.
To supply to this lack, numerous analytical representations (called equations of state) have been suggested~\cite{Davis1967, MacDonald1969}, which allows to extrapolate the density measurements carried out at moderate pressures. However, different methods often lead to incompatible results~\cite{Sun2006b} and the effectiveness for such analytical predictions needs to be validated against experimental high pressure data. 

In this work, we report the measurements of sound velocity in liquid mercury at high pressures obtained by the picosecond acoustics technique~\cite{Thomsen1986a,Decremps2008, Chigarev2008, Wright2008, Decremps2009} 
coupled with a surface imaging technique~\cite{Sugawara2002, Decremps2010, Zhang2011}.

\section{Experimental set-up}

\begin{figure}[ht]
  \centering
  \includegraphics[width=\linewidth]{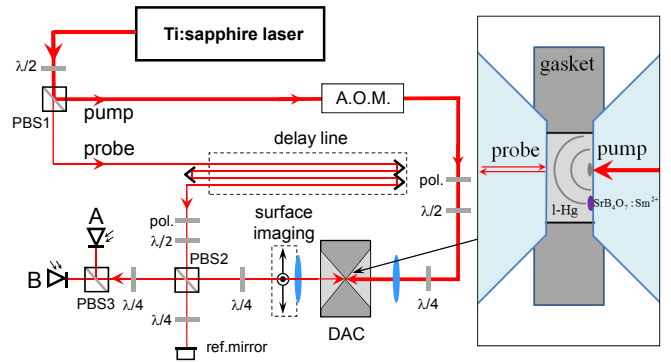}
  \caption{\label{fig:schema_experimental}Schematic set-up of the picosecond acoustics experiment involving a surface imaging set-up and a Michelson interferometer.
  The inset shows the sample in the DAC with the pump and probe on the opposite sides.
  PBS~: polarizing beam-splitter, $\lambda$/4~: quarter wave plate, $\lambda$/2~: half-wave plate, pol.~: linear polarizer, A.O.M.~: acousto-optic modulator. See text for the details.}
\end{figure}

The experimental set-up used in this work is shown in the figure~\ref{fig:schema_experimental}.
The light source is a femtosecond Ti:sapphire laser delivering $\lambda$=800~nm light pulses of about 100~fs width, at a repetition rate of 79.66~MHz.
The output of this pulse laser is splitted into pump and probe beams. The pump beam is modulated at the frequency of 1 MHz by an acousto-optic modulator.
A lock-in amplifier synchronized with the modulation frequency is used to improve the signal-to-noise ratio. The pump is focused onto a small spot (3 $\mu$m) of the sample which creates a sudden and small temperature rise of about 1 K. The corresponding thermal stress generated by thermal expansion relaxes by launching a longitudinal acoustic strain field mainly along the direction perpendicular to the flat parallel faces of the sample. The probe beam is delayed according to the pump through a 1~m length delay line used in 4~pass allowing a total temporal range of 13.333~ns with 1~ps step. Probe beam is focused into a spot on the opposite surface of the sample. 
The variation of reflectivity as a function of time is detected through the probe intensity variations. 
These variations are due to thermal and acoustic effects which alter the optical reflectivity. The photo-elastic and the photo-thermal coefficients contribute to the change of both the imaginary and real part changes of the reflectivity\cite{Duquesne2003}, whereas the surface displacement only modifies the imaginary part. The detection is carried out by a stabilized Michelson interferometer which allows the determination of both the reflectivity imaginary and real part changes\cite{Perrin1999}.
A 100~$\mu$m~x~100~$\mu$m surface imaging of the sample can be done by a scan of the probe objective mounted on a 2D translational motor.

A membrane diamond anvil cell (DAC) is used as high pressure generator. The pressure is determined by the shift of the SrB$_{4}$O$_{7}$:5\%Sm$^{2+}$ fluorescence line which is known to be temperature independent ~\cite{Datchi1997} with an accuracy of 0.1 GPa. To reach high temperatures the DAC is placed in a resistive furnace.
The temperature measurements were calibrated with the well known melting line of Hg\cite{Klement1963} and checked by a thermocouple glued on the diamond. The relative uncertainty on the temperature is estimated around 1$\%$.

Ultra-pure mercury (99.99 $\%$) from Alfa Aesar was used during the whole sets of experiments.
The gasket material was rhenium, known to be chemically inert at high temperature with mercury\cite{Guminski2002}. Moreover, no reaction is expected between carbon (i.e. diamonds) and mercury~\cite{Guminski1993}. A small droplet of liquid mercury was loaded in a 200 $\mu$m~diameter gasket hole whose thickness is between 20 and 70~$\mu$m. A large diameter hole is here chosen to avoid acoustic reflections from the edges of the gasket.

\section{Sound velocity measurements}

\subsection{Temporal method}

\begin{figure}
\includegraphics[width=.45\linewidth]{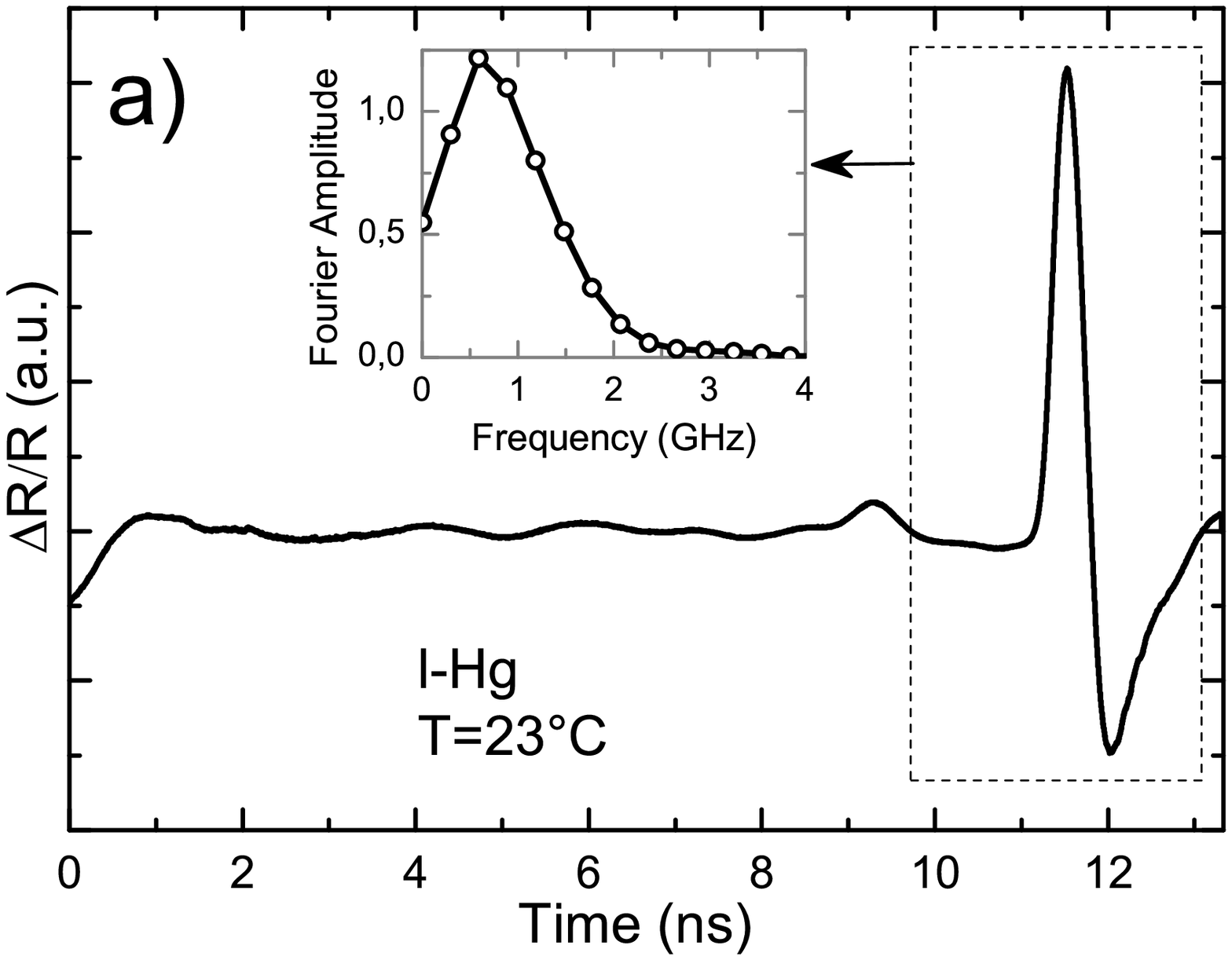}
\includegraphics[width=.462\linewidth]{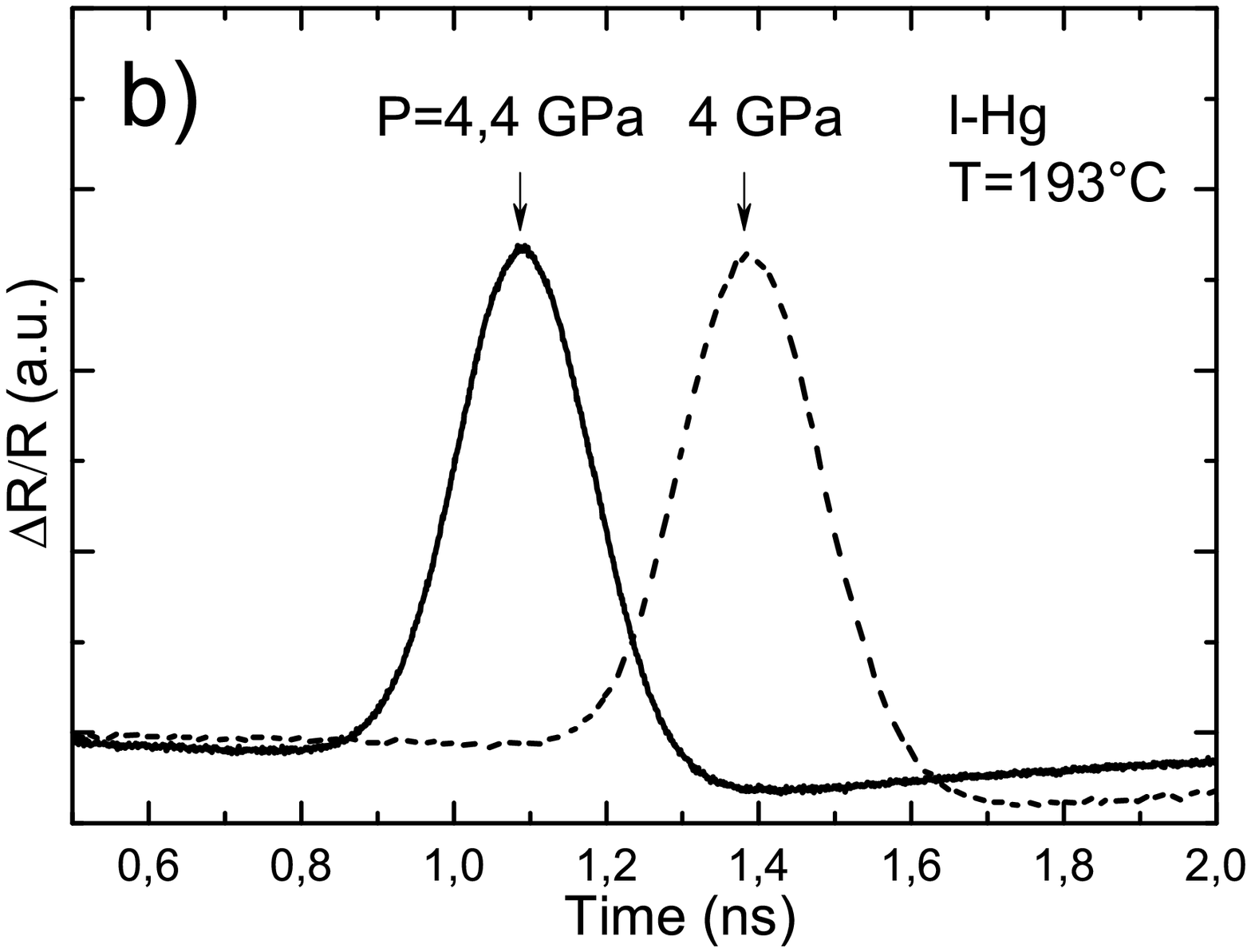}
\caption{(a) Variation of reflectivity as a function of time with collinear pump and probe beams focused on the opposite sides of the sample. 
The signal inside the dashed box corresponds to the main acoustic wavefront arriving at the diamond/mercury interface.
(inset) Fourier transform of this signal.
  (b) The peak position shift is mainly due to the variation of sound velocity with pressure.}\label{fig:variation_temporels}
\end{figure}

In this first configuration, the probe beam is focused to a spot on the opposite surface of the sample with respect to the surface illuminated by the pump, the two beams being collinear. 
The variation of reflectivity as a function of time is detected through the measurement of the intensity modification of the probe, which is delayed with respect to the pump by a different optical path length. After propagation along the sample, the contribution of acoustic effects alter the optical reflectivity of the opposite surface of the sample opposite to the incoming beam. As a matter of fact, a peak in the reflectivity is observed as soon as the acoustic waves reach the sample surface (see figure~\ref{fig:variation_temporels}a). Note that, since liquid mercury is fully embedded into the gasket hole, the peak-echo arises at a time $\Delta t$ corresponding to a single way of the acoustic wave into mercury along the thickness between the surfaces of the two DAC diamond culets.

In the present study, for each pressure and temperature condition, longitudinal acoustic echoes have been systematically observed in the recorded time variation of the reflectance imaginary part. 
For two different pressures at a fixed temperature, the figure~\ref{fig:variation_temporels}b) illustrates the temporal shift of the acoustic echo.

Along four isothermal experiments, the sound velocity $v$ as a function of pressure was derived using this method (here called "temporal" method) from the pressure variations of $\Delta t$ and from the use of the following equation~(\ref{eq:thickness})~\footnote{In addition, the shift of the temporal echo allows to measure accurately the melting line for nontransparent materials~\cite{Decremps2009}.}:

\begin{equation}
  e_0 = v \Delta t = v(t_{0}+pT_{laser}-\tau)
  \label{eq:thickness}
\end{equation}

where  $e_0$ is the gasket thickness (note that here, the subscripted "0" indicates that the measurement is carried out with a particular and fixed spatial position of the probe, i.e. in the axis of the pump beam). $t_{0}$ is the corresponding emergence time of the wave at the surface of the diamond culet as observed on the recorded reflectivity variation. In order to relate $t_{0}$ to the relevant travel time $\Delta t$, it is here required to determine $p$ and $\tau$. The integer $p$ takes into account the successive generation of echoes due to the the laser repetition rate, and the value of $\tau$ corresponds to the time at which the pump-probe coincidence occurs. $p$ is first "hand-made" assigned taking into account a rough estimated value of the travel time. We will explain in the next subsection how the validation of the $p$ value can be done using the imagery method. Concerning $\tau$, we have previously measured it in an aluminum thin film outside and inside the DAC (the variation of the optical path due to the presence of diamonds DAC, around 2~mm thick, is negligible). For the present set-up, we obtained $\tau = 0.330 \pm 0.002$~ns.

Whereas this configuration allows a simple and quick way to extract the sound velocity, its major and evident disadvantage comes from the need to know the gasket thickness at each thermodynamical conditions.
We thus have developed an additional set-up, called "imagery" method, in order to be able to determine both $v$ and $e_0$ experimental values.

\subsection{Imagery method}

\begin{figure}[ht]
  \includegraphics[width=0.4\linewidth]{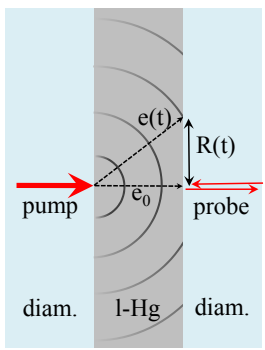}
  \caption{Schematic diagram of the liquid mercury inside the DAC (note that diam. stands for diamond anvil). Spherical wavefronts produced by the tightly focused pump beam are due to acoustic diffraction. On the probe side, the radius $R(t)$ of the circles appearing on the surface depends on the sample thickness $e_{0}$ and on the radius of the acoustic wavefront $e(t)$.}
  \label{fig:spherical_waves}
\end{figure}

For a given pump-probe delay the bulk spherical wavefront generated by the pump laser reaches the opposite surface of the sample and produces circular patterns.
The spherical wavefront is due to acoustics diffraction produced by the tightly focused laser beam as described in figure~\ref{fig:spherical_waves}.
In the source near-field the corresponding wavefront is complex but it can be demonstrated that the detection occurs on the opposite side of the sample, i.e. in the far-field.
The transition between the near-field and the far field occurs at the depth~\cite{Lin1990} :
\begin{equation}
  z_{t} \approx \frac{d^{2}}{4\lambda_{ac}}
\end{equation}
where d=3~$\mu$m is the diameter of the laser spot and $\lambda_{ac}$ the mean wavelength of the acoustic wave packet.
The mean frequency of the wave packet extracted from a Fourier transform of a temporal scan is roughly 0.6~GHz (see the inset of figure~\ref{fig:variation_temporels}a) leading to $\lambda_{ac}\approx 1 \mu m$. 
The transition distance $z_{t}=2~\mu m$ is thus well lower than the sample thickness which simplifies the analysis of the detection process, done in the far field approximation.

\begin{figure}[ht]
  \includegraphics[width=7cm]{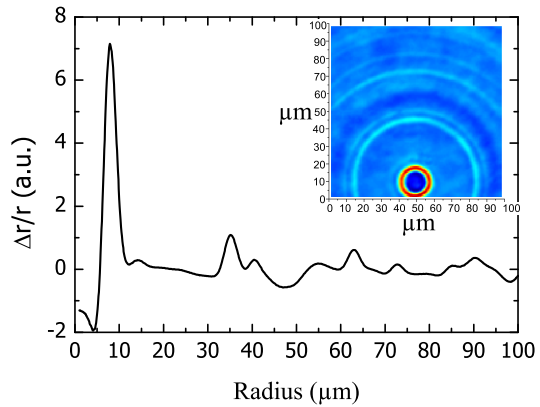}
  \caption{\label{fig:pattern_analyse} Change in the reflectivity of liquid mercury at 1~GPa and 30\char23C as a function of the optical probe-pulse time delay. Inset : corresponding experimental image of acoustic wavefront on the surface of liquid Hg with an arbitrary color scale, and associated profile integration in liquid mercury at delay t=5.3~ns.}
\end{figure}

A typical 100~$\mu m$ x 100~$\mu m$ image associated with the integrated intensity profile is shown in figure~\ref{fig:pattern_analyse}.
The center of the acoustics rings is spatially determined with an uncertainty lower than $\pm 0.5~\mu m$ in the two directions perpendicular to the beam.
Due to the repetition rate of the laser, theses patterns are renewed every $T_{laser}$=12.554~ns.
Perfect circular rings are expected in the liquid phase (taking into account that diamond culets stays parallels and are not deformed at high pressure~\cite{Hemley1997}).
The solid phase of mercury ($\alpha$-Hg) is easily detected since the acoustics pattern is here no longer circular due to the anisotropy of the crystal.

\begin{figure}[ht]
  \includegraphics[width=\linewidth]{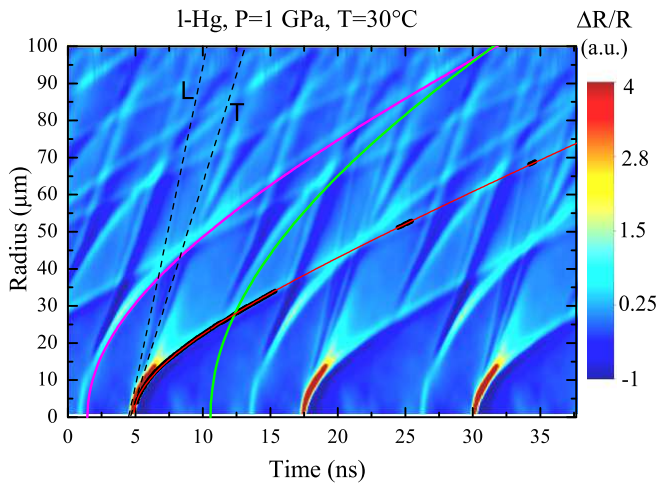}
  \caption{\label{fig:time_radius}Integrated profiles as a function of time at 1~GPa and 30\char23 C.
  The time scale is extended by repetition of the picture each $T_{laser}$.
  The evident ripples in this picture sign the growing of circular wavefronts on the surface of the sample.
  This pictures show the principal wavefront (red line), first and second reflections (pink and green lines respectively), and longitudinal and transverse SSBW in diamond (dashed lines).
  The experimental points are represented by black circles.
  Red line is the fit with eq.~(\ref{eq:radius_with_time}) whereas pink and green lines correspond to calculation with eq.~(\ref{eq:radius_with_time_n}), as discussed in the main text.
}
\end{figure}

For each thermodynamical condition, the acoustic wave front image is recorded as a function of pump-probe delay, with a time step of 0.1~ns.
All the corresponding integrated profiles can be stacked together into a graph (fig.~\ref{fig:time_radius}) where the vertical color scale indicates the regions of high (red) and low (blue) reflectivity.
The main longitudinal wave propagating in mercury appears at 5, 17.5 and 30~ns. The waves at 1, 13.5, 26~ns and 11, 23.5, 36~ns correspond to the first and second reflections of the main bulk wave, respectively.
As evident by (fig.~\ref{fig:time_radius}) some ripples have a linear dependance of the radius with time. This is the signature of surface skimming bulk waves (SSBW) propagating in the diamond parallel to the surface \footnote{These waves arise at the critical angle of the Snell-Descartes law of acoustic refraction at the diamond-mercury interface. Above this critical angle there is total reflection and any other SSBW cannot be generated. This angle is estimated at 4.3\char23 between the propagating wave vector of the mercury bulk wave and the diamond surface. This weak angle imposes the SSBW are generated roughly at $t_0$. In interferometry two kinds of SSBW are visible with mean velocities of $c_{T}=12\pm 1$~km/s and $c_{L}=18 \pm 1$~km/s corresponding to the transverse and longitudinal velocities in the diamond as expected.}.

Considering the evolution of spherical wavefronts inside the sample as shown in (figure~\ref{fig:spherical_waves}), the time evolution of the ring diameters $R(t)$ is given by :

\begin{equation}
  R(t)=\sqrt{e^2(t)-e_0^2}
  \label{eq:radius_with_time}
\end{equation}
where $e(t)=v(pT_{laser}+t-\tau)$ is the distance covered by the wavefront acoustics wave inside the sample.

Eq.~(\ref{eq:radius_with_time}) is fitted to the experimental radius (black dots on fig.~\ref{fig:time_radius}) with the sound velocity $v$ and the arrival time $t_{0}$ as free parameters. 
Then the thickness $e_{0}$ is deduced with eq.~\ref{eq:thickness}.
Knowing the velocity $v$ and the sample thickness $e_{0}$, it is now straightforward to predict the radius evolution $R_{n}(t)$ for the $n^{th}$ reflected wave 

\begin{equation}
   R_{n}(t)=v\sqrt{(t+p_{n}T_{laser}-\tau)^{2}-(t_{0}^{n}+p_{n}T_{laser}-\tau)^{2}}
   \label{eq:radius_with_time_n}
\end{equation}
where $p_{n}$ is an integer properly chosen and $t_{0}^{n}$ can be calculated from $t_{0}$
\begin{equation}
  t_{0}^{n}=(2n+1)t_{0}-2n\tau+T_{laser}[(2n+1)p-p_{n}].
  \label{eq:reflection}
\end{equation}
Pink and green lines shown in fig.~\ref{fig:time_radius} stand for the first and the second reflections respectively.  
The good agreement between the ripples and the lines confirm the assignment previously done.

An accurate determination of the experimental radius is related to a correct interpretation of the integrated profile, to avoid systematic errors on the parameters $t_{0}$ and $v$.
The integrated profile shows an antisymetric wavelet (as observed in the figure~\ref{fig:pattern_analyse} between 0 and 10~$\mu m$), related to the bipolar strain of the acoustic pulse
~\footnote{The bipolar profile can be explained by the generation, propagation and detection process involving the acoustic pulse. 
The exact theory is beyond the scope of this paper however let us give a brief explanation.
The thermoelastic generation in the liquid Hg bonded to the diamond produces an unipolar and asymmetric strain profile~\cite{Wright2008}.
During its propagation from the near field to the far field, the acoustic pulse transforms from unipolar to bipolar shape.
This transformation is explained by the Gouy phase shift due to the acoustic diffraction~\cite{Holme2003}.
Finally the shape of the echo in the integrated profile is roughly related to the shape of the pulse~\cite{Wright2008}.}. 
Once generated, the acoustic pulse is immediately reflected at the interface diamond/mercury. 
As a consequence, the spatial extension of the pulse is doubled~\cite{Wright2008}. The part of the pulse generated exactly at $t=\tau$ is the midpoint of this pulse. 
Thus the experimental value of the radius corresponds to the midpoint of the perturbation seen in surface, in this case the inflection point. 
Finally, the alteration of the integrated profile due to the acoustic dispersion is supposed negligible~\cite{Wright2008}.

\subsection{Results}

\begin{figure}[h]
  \includegraphics[width=\linewidth]{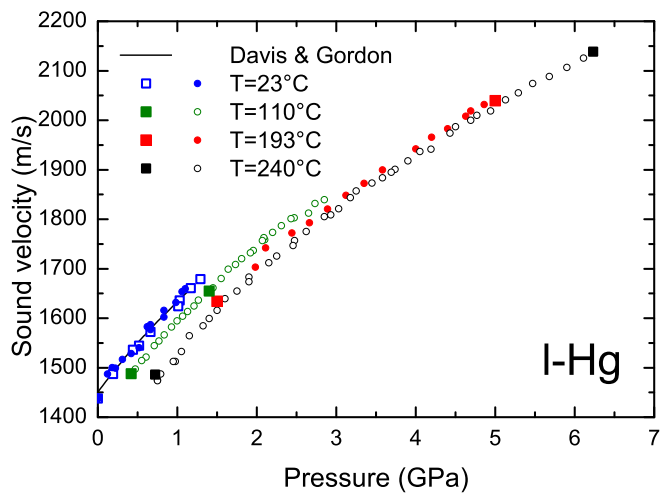}
  \caption{\label{fig:final_velocities} Adiabatic sound velocity in liquid mercury as a function of pressure at various temperatures. Squares : imagery method. Circles : temporal method. The line is from Davis~\cite{Davis1967}.}
\end{figure}

\begin{figure}[h]
  \includegraphics[width=\linewidth]{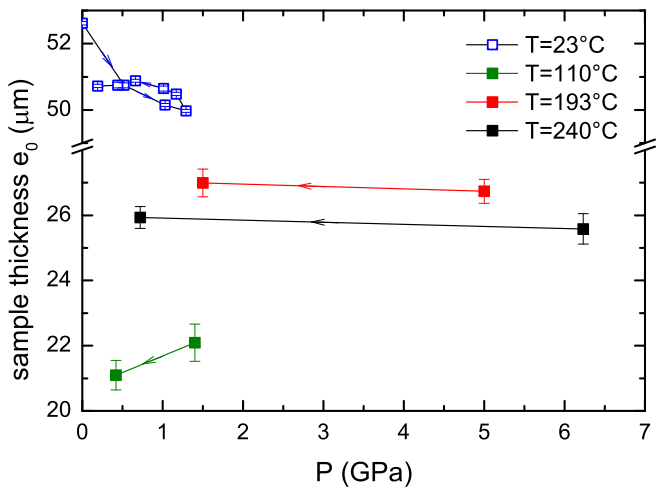}
  \caption{\label{fig:epaisseurs} Sample tichnesses as a function of pressure for each isotherm. The arrows indicate the succession of the measurements.}
\end{figure}

We however emphasize that, while this imaging configuration is thus very powerful (both thickness and sound velocity of the sample are determined using a self consistent method), it 
has the main disadvantage to be very time consuming nothing compare to the couple of seconds needed by the "temporal method".
We thus only used the imagery method at few pressures (about three or four pressures per isotherms) in order to extract both $v$ and $e_{0}$ for each point.  
The velocities and the corresponding thicknesses obtained by the imagery method are shown respectively in the fig.~(\ref{fig:final_velocities}) and the fig.~(\ref{fig:epaisseurs}) by the squares. 
At ambient condition, the sound velocity is $1450~\pm~15$~m/s in good agreement with the previous studies~\cite{Tilford1987,Davis1967}. 
Upon the pressure downstroke we observed a weak pressure dependence of the thickness, as previously published~\cite{Dewaele2003}. 
Although the volume increases when the pressure decrease, the thickness remains constant due to complex plasticity process inside the gasket~\cite{Dunstan1989}. 
A simple linear interpolation of these experimental points is used and provides a reliable estimation of the thickness variation as a function of pressure for the whole pressure and temperature range of the experiments.
The sample thickness being known, the sound velocity can be directly extracted by a scan with temporal method. 
Figure~\ref{fig:final_velocities} summarizes our complete results in pressure up to 7~GPa and temperature up to 240\char23 C. 
The experimental data set is reported in the table~\ref{tab:tablevelocities}.
The velocities obtained at ambient temperature agree with the data from Davis~\cite{Davis1967} up to 1.2~GPa. 

\begin{table}[h]
\caption{\label{tab:tablevelocities} Sound velocity in liquid Hg. Values marked by a star (*) are obtained by the imagery method.}
\begin{ruledtabular}
\begin{tabular}{c  c  c  c  c  c  c  c}
\multicolumn{2}{c}{T=23 \char23 C} & \multicolumn{2}{c}{T=110\char23C} & \multicolumn{2}{c}{T=193\char23C} & \multicolumn{2}{c}{T=240\char23C}\\
P & v & P & v & P & v & P & v \\
(GPa) & (m/s) & (GPa) & (m/s) &(GPa) & (m/s)  & (GPa) & (m/s)  \\
\hline
0.00*	&	1438	&	0.42*	&	1488	&	1.50*	&	1634	&	0.72*	&	1486	\\
0.00		&	1442	&	0.44		&	1485	&	1.51		&	1633	&	0.75		&	1474	\\
0.12		&	1487	&	0.47		&	1498	&	1.51		&	1635	&	0.75		&	1475	\\
0.18		&	1500	&	0.55		&	1514	&	1.98		&	1703	&	0.79		&	1487	\\
0.19*	&	1487	&	0.60		&	1521	&	2.11		&	1742	&	0.95		&	1513	\\
0.22		&	1499	&	0.71		&	1545	&	2.44		&	1772	&	0.97		&	1513	\\
0.31		&	1517	&	0.77		&	1554	&	2.66		&	1793	&	1.05		&	1533	\\
0.42		&	1529	&	0.83		&	1567	&	2.89		&	1821	&	1.15		&	1565	\\
0.44*	&	1536	&	0.93		&	1583	&	3.12		&	1848	&	1.33		&	1584	\\
0.52		&	1541	&	1.00		&	1595	&	3.35		&	1873	&	1.40		&	1600	\\
0.52*	&	1544	&	1.06		&	1604	&	3.58		&	1900	&	1.50		&	1616	\\
0.62		&	1583	&	1.13		&	1614	&	4.00		&	1942	&	1.60		&	1639	\\
0.66		&	1587	&	1.21		&	1625	&	4.20		&	1966	&	1.75		&	1655	\\
0.66*	&	1572	&	1.27		&	1637	&	4.40		&	1983	&	1.90		&	1674	\\
0.66		&	1577	&	1.39		&	1657	&	4.63		&	2008	&	1.90		&	1683	\\
0.83		&	1603	&	1.40		&	1655	&	4.69		&	2019	&	2.15		&	1712	\\
0.83		&	1616	&	1.40*	&	1655	&	4.86		&	2032	&	2.25		&	1726	\\
0.98		&	1632	&	1.45		&	1662	&	5.00*	&	2039	&	2.46		&	1747	\\
1.01*	&	1624	&	1.55		&	1680	&	--		&	--	&	2.48		&	1757	\\
1.03*	&	1636	&	1.64		&	1699	&	--		&	--	&	2.62		&	1775	\\
1.06		&	1654	&	1.73		&	1708	&	--		&	--	&	2.85		&	1805	\\
1.10		&	1659	&	1.81		&	1720	&	--		&	--	&	2.93		&	1809	\\
1.17		&	1663	&	1.92		&	1732	&	--		&	--	&	3.03		&	1821	\\
1.17*	&	1661	&	1.96		&	1737	&	--		&	--	&	3.17		&	1844	\\
--		&	--	&	2.07		&	1757	&	--		&	--	&	3.25		&	1857	\\
--		&	--	&	2.09		&	1763	&	--		&	--	&	3.45		&	1873	\\
--		&	--	&	2.10		&	1758	&	--		&	--	&	3.58		&	1884	\\
--	&	--	&	2.20		&	1773	&	--		&	--	&	3.69		&	1895	\\
--		&	--	&	2.31		&	1787	&	--		&	--	&	3.74		&	1901	\\
--		&	--	&	2.43		&	1801	&	--		&	--	&	3.90		&	1918	\\
--		&	--	&	2.47		&	1803	&	--		&	--	&	4.05		&	1937	\\
--		&	--	&	2.65		&	1812	&	--		&	--	&	4.19		&	1942	\\
--		&	--	&	2.73		&	1832	&	--		&	--	&	4.43		&	1974	\\
--		&	--	&	2.85		&	1839	&	--		&	--	&	4.50		&	1987	\\
--		&	--	&	--		&	--	&	--		&	--	&	4.69		&	2000	\\
--		&	--	&	--		&	--	&	--		&	--	&	4.77		&	2010	\\
--		&	--	&	--		&	--	&	--		&	--	&	4.94		&	2019	\\
--		&	--	&	--		&	--	&	--		&	--	&	5.13		&	2041	\\
--		&	--	&	--		&	--	&	--		&	--	&	5.29		&	2056	\\
--		&	--	&	--		&	--	&	--		&	--	&	5.47		&	2074	\\
--		&	--	&	--		&	--	&	--		&	--	&	5.68		&	2089	\\
--		&	--	&	--		&	--	&	--		&	--	&	5.90		&	2107	\\
--		&	--	&	--		&	--	&	--		&	--	&	6.11		&	2126	\\
--		&	--	&	--		&	--	&	--		&	--	&	6.23		&	2139	\\
--		&	--	&	--		&	--	&	--		&	--	&	6.23*	&	2139	\\
\end{tabular}
\end{ruledtabular}
\end{table}

\section{Compression curve}
\label{sec:eos}

\subsection{Thermodynamical relations}

The density variations as a function of pressure and temperature can be extracted from the sound velocity measurements via classical thermodynamic relations~\cite{Davis1967,Daridon1998,Lago2008,Davila2009}.
The adiabatic sound velocity~\footnote{The sound waves propagate adiabatically up to a frequency $f$ given by $f=v^2 \rho C_V / 2 \pi \kappa$ where $\kappa$ is the thermal conductivity and $C_{V}$ the isochoric specific heat~\cite{Fletcher1974}. In the liquid mercury $f\approx$100~GHz well above the 10~GHz reached in our experiments.} $v$ is related to the adiabatic compressibility by $\beta_{S}=1/\rho v^{2}$ and to the thermal compressibility $\beta_T=1/\rho (\partial \rho/\partial P )_T$ by
\begin{equation}
  \beta_{T}=\beta_{S}+\frac{T\alpha_{P}^2}{\rho C_P}
  \label{eq:adcompress1}
\end{equation}
where $C_P$ is the isobaric heat capacity and $\alpha_{P}$ is the thermal expansion coefficient at constant pressure defined by
\begin{equation}
  \alpha_{P}\equiv-\frac{1}{\rho}\left(\frac{\partial \rho}{\partial T}\right)_{P}.
  \label{eq:defalpha}
\end{equation}
Relation~\ref{eq:adcompress1} can be rewritten as
\begin{equation}
 \left(\frac{\partial \rho}{\partial P}\right)_T = \frac{1}{v^{2}}+\frac{T\alpha_{P}^2}{C_P}.
  \label{eq:adcompress}
\end{equation}
The integration of equation~(\ref{eq:adcompress}) between arbitrary pressures $P_1$ and $P_2$ leads to the equation
\begin{equation}
  \rho(P_2,T)-\rho (P_1,T)=\int_{P_1}^{P_2} \frac{dP}{v^{2}(P,T)}+T \int_{P_1}^{P_2}\frac{\alpha_{P}^{2}(P,T)}{C_P(P,T)} dP
  \label{eq:density}
\end{equation}
where the variation of $C_P$ with pressure can be evaluated via
\begin{equation}
   \left(\frac{\partial C_P}{\partial P}\right)_{T}= - \frac{T}{\rho} \left\{   \left(\frac{\partial \alpha_P}{\partial T}\right)_{P}+\alpha_{P}^2\right\}.
   \label{eq:CPsurdP}
\end{equation}
The three equations~(\ref{eq:defalpha}), ~(\ref{eq:density}) and~(\ref{eq:CPsurdP}) are used into the modified recursive procedure~\cite{Daridon1998} described in the following in order to obtain the density as a function of pressure and temperature.

\subsection{Recursive numerical procedure}

This procedure needs as input parameters the sound velocity as a function of temperature and pressure, and the temperature variations of $\rho$ and $C_P$ at room pressure $P_0$.
All high pressure sound velocity values come from our work (see table~(\ref{tab:tablevelocities}) and figure~\ref{fig:final_velocities}). 
At ambient pressure the data from Coppens \textit{et al}\cite{Coppens1967} and Jarzynski~\cite{Jarzynski1963} $v(P_{0},T)$ are appended to our experimental values. 
All data are interpolated and smoothed by a polynomial function $P=\sum^{}_{i,j} a_{ij}T^{i}v^{j}$.
The function $P(T,v)$ is chosen because it predicts a steady increase of velocity with pressure, whereas the function $v(P,T)$ leads to a maximum in velocity as a function of pressure without any physical meaning~\cite{Davis1967}.
The coefficients $a_{ij}$ are shown in table~(\ref{tab:Coefvitesse}).

\begin{table}[h]
\caption{\label{tab:Coefvitesse} $a_{ij}$ coefficients for $P=\sum_{i,j} a_{ij}T^{i}v^{j}$ with $P$ in GPa and T in Celsius and $v$ in m/s.}
\begin{ruledtabular}
\begin{tabular}{ccccccc}
i/j	& 0                 & 1                 & 2	\\
\hline
0	& $4.00423$	        & $-1.006 \cdot 10^{-2}$	& $5.01139 \cdot 10^{-6}$\\
1	& $5.81263 \cdot 10^{-4}$	& $1.23466 \cdot 10^{-6}$	& -- \\
2	& $8.17121 \cdot 10^{-7}$	& --                  & -- \\
\end{tabular}
\end{ruledtabular}
\end{table}

The density $\rho(P_0,T)$ is calculated from the polynomial formula given by Holman (equation (28) in the Ref.~\cite{Holman1994}).
The coefficient of thermal expansion $\alpha_{P}(P_0,T)$ is directly deduced from the density using equation~(\ref{eq:defalpha}).
The values of heat capacity $C_{P}(P_0,T)$ between 273~K and 800~K are interpolated from the measured values of Ref.\cite{Douglas1951} using a third order polynomial function. The best interpolation relation obtained is $C_{P}(P_0,T)=152.77098-0.06585~T+6.89659 \cdot 10^{-5}~T^{2}-1.29761 \cdot 10^{-8}~T^{3}$ with T in K and $C_P$ in $J\cdot kg^{-1} \cdot K^{-1}$. 

Starting from the room pressure values $P_0$, the values at higher pressures are obtained by a small pressure increment $\Delta P=P_2-P_1=0.01$~GPa from the already determined pressure step $P_1$ to the next calculated pressure point $P_2$.
At each pressure step, the quantities $\rho$, $\alpha_{P}$ and $C_{P}$ are calculated.
All the quantities are evaluated between 20 and 240\char23C with a temperature step $\Delta T=1$~K. 
The main expression is the equation~(\ref{eq:density}) involving the density calculation at $P_{2}$. 
The first integral in the equation~\ref{eq:density} is evaluated numerically with the function $P(v,T)$.
This term represent the major contribution of the variation of density and it depends only on velocity and an accurate numerical integration.
The second term in the equation~(\ref{eq:density}) contributes for roughly 15\% of the density value and is evaluated iteratively until convergence.
In the first step of the iterative process the quantity $T \alpha_{P}^{2}/C_P$ is kept constant leading to a first crude approximation of the density at pressure $P_2$.
Then $\alpha(P_2,T)$ is deduced from this first approximation using equation~(\ref{eq:defalpha}).
In the second iterative step the variation of $\alpha_{P}$ is taken into account by a linear interpolation between $\alpha_P(P_1,T)$ and $\alpha_P(P_2,T)$ and introduced in equation~(\ref{eq:density}) while $C_{P}$ is still kept constant leading to a first refinement of the density value.
This process is repeated until the convergence of the $\rho(P_2,T)$ value is reached.
During this procedure, $\alpha_{P}(T)$ is smoothed by a third order polynom to avoid the side effects occurring with the numerical derivation.
Finally, the $C_P(P_2,T)$ value at the new pressure $P_2$ is obtained by a linear extrapolation of the $C_P(P_1,T)$ value at $P_1$ and its derivative through equation~(\ref{eq:CPsurdP}).

In order to evaluate the robustness of this numerical procedure, we have performed a test on the well known thermodynamic data of liquid water~\cite{Wagner2002} in the temperature range 280-340~K and the pressure range 0.1-50~MPa with $\Delta T=1$~K and $\Delta P=$0.1~MPa.
We have obtained results in very good agreement with the literature data, 
the comparison providing the relative uncertainties due to the numerical procedure.
The uncertainties are $\pm 0.002 \%$ for density, $\pm 1 \%$ for thermal expansion and $\pm 0.5 \%$ for heat capacity.

\subsection{Results}

\begin{figure}[h]
  \begin{center}
       \includegraphics[width=\linewidth]{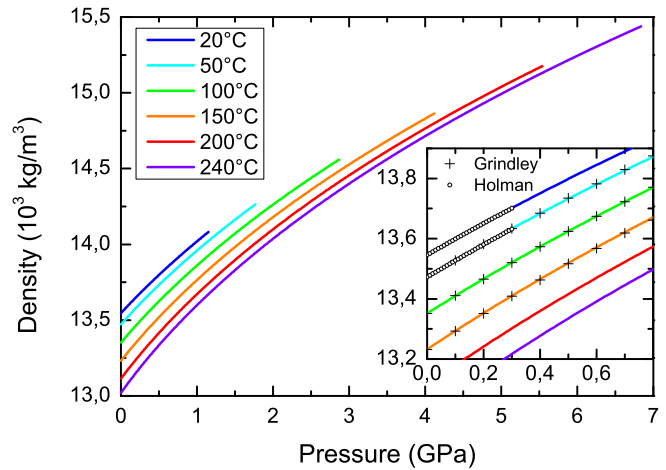}
  \end{center}
  \caption{\label{fig:rho_final} Density $\rho$ of liquid Hg versus pressure at different temperatures.
  The calculation procedure from sound velocity measurements is explained in the text. Inset : Comparison between our work and the data from Holman~\cite{Holman1994} and Grindley~\cite{Grindley1971}.}
\end{figure}

\begin{figure}[h]
  \includegraphics[width=\linewidth]{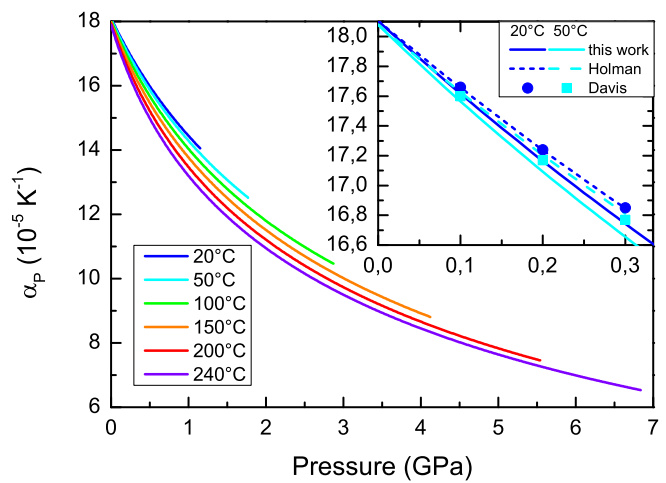}
  \caption{\label{fig:alpha_final} The coefficient of thermal expansion $\alpha_{P}$ of liquid Hg as a function of pressure at different temperatures.
  (inset) Comparison with the data of Davis~\cite{Davis1967} and Holman~\cite{Holman1994}.}
\end{figure}

\begin{figure}[h]
  \includegraphics[width=\linewidth]{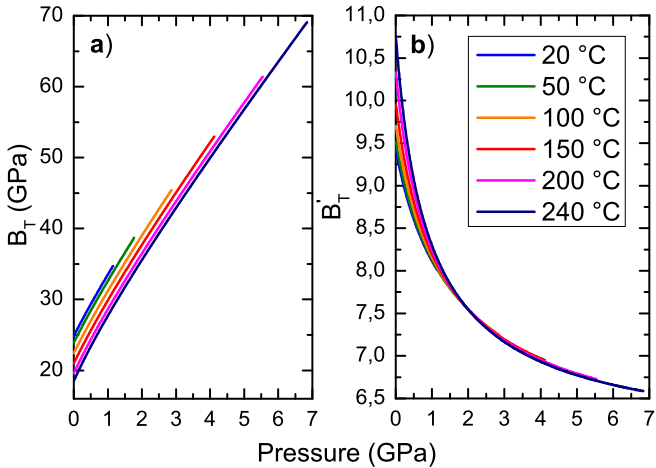}
  \caption{\label{fig:BT_final}a)~Isothermal bulk modulus $B_T$ as a function of pressure at different temperatures. b)~First derivative of the isothermal bulk modulus $B_T^{'}$ as a function of pressure at different temperatures.}
\end{figure}

\begin{figure}[h]
  \includegraphics[width=\linewidth]{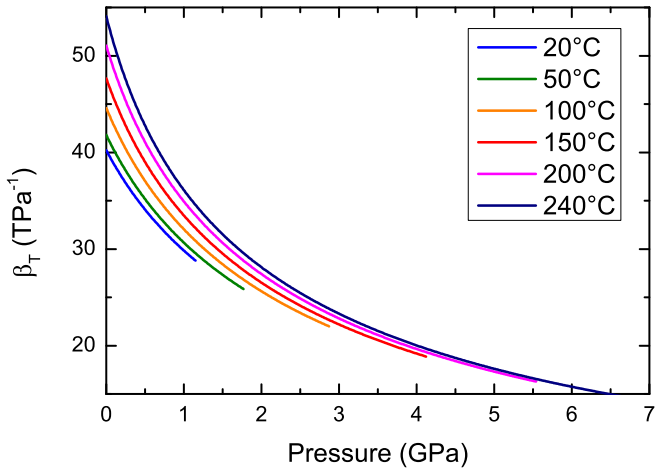}
  \caption{\label{fig:betaT_final}Isothermal compressibility $\beta_T$ as a function of pressure at different temperatures.}
\end{figure}

The quantities $\rho$ and $\alpha_P$ as a function of temperature and pressure up to the melting line are shown in figures~\ref{fig:rho_final} and~\ref{fig:alpha_final}, respectively. 
The values of the density are reported in the table~(\ref{tab:tabrhovalues}).
The derived quantities are the isothermal compressibility $\beta_{T}$, isothermal bulk modulus $B_T=\beta_{T}^{-1}$ and the first derivative of the bulk modulus $B_T^{'}=(\partial B_T / \partial P)_{T}$. 
They are shown in the figures \ref{fig:betaT_final} and \ref{fig:BT_final} respectively.

The uncertainties on final parameters of liquid mercury have been evaluated by the introduction of small perturbations in the three input quantities $v(P,T)$, $\rho(P_0,T)$ and $C_{P}(P_0,T)$.
An increase or decrease of the sound velocities data by 10~m/s leads to a variation of $\pm 0.15 \%$ of the density, $\pm 1.7 \%$ of the thermal expansion and $\pm 0.7 \%$ of the heat capacity.
The relative uncertainty in $\rho(P_0)$~\cite{Holman1994} is roughly $10^{-6}$ and accounts for a $\pm 0.5 \%$ relative variation of the thermal expansion and $\pm 0.8 \%$ of the heat capacity.
According to Douglas~\cite{Douglas1951} the heat capacity $C_{P}(P_0,T)$ is known at $\pm 0.3 \%$. This leads to a $\pm 1.2 \%$ relative variation of the thermal expansion and $\pm 1 \%$ of the final heat capacity.
Finally, the different uncertainties are quadratically summed. The maximal uncertainties associated to the absolute measurements of the different quantities are around $\pm 0.15 \%$ for the density, $\pm 3.8 \%$ for the thermal expansion and $\pm 3.4 \%$ for the heat capacity. 
The quantity $C_P$ is not shown because the variations of $C_{P}$ deduced from the numerical procedure have the same order of magnitude than the uncertainty.

\begin{table}\footnotesize 
 \caption{Density in liquid Hg (kg/m$^{3}$) calculated by the procedure explained in the section~(\ref{sec:eos}).} 
\begin{ruledtabular}
\begin{tabular}{lcccccc}
P(GPa) & \multicolumn{6}{c}{T (\char23 C)}   \\
 &  20 &  50 &  100 &  150  &  200 &  240 \\ 
  \hline
0.0	&	13546	&	13473	&	13351	&	13232	&	13113	&	13020	\\
0.2	&	13651	&	13581	&	13466	&	13353	&	13242	&	13154	\\
0.4	&	13751	&	13684	&	13574	&	13466	&	13360	&	13277	\\
0.6	&	13845	&	13780	&	13674	&	13571	&	13470	&	13391	\\
0.8	&	13934	&	13872	&	13770	&	13671	&	13574	&	13498	\\
1.0	&	14020	&	13960	&	13861	&	13765	&	13672	&	13599	\\
1.2	&	--	&	14043	&	13948	&	13855	&	13765	&	13695	\\
1.4	&	--	&	14124	&	14031	&	13941	&	13854	&	13786	\\
1.6	&	--	&	14201	&	14111	&	14024	&	13939	&	13873	\\
1.8	&	--	&	--	&	14188	&	14103	&	14021	&	13957	\\
2.0	&	--	&	--	&	14263	&	14180	&	14099	&	14037	\\
2.2	&	--	&	--	&	14334	&	14254	&	14175	&	14115	\\
2.4	&	--	&	--	&	14404	&	14325	&	14249	&	14189	\\
2.6	&	--	&	--	&	14472	&	14394	&	14320	&	14262	\\
2.8	&	--	&	--	&	14537	&	14462	&	14389	&	14332	\\
3.0	&	--	&	--	&	--	&	14527	&	14456	&	14400	\\
3.2	&	--	&	--	&	--	&	14591	&	14521	&	14466	\\
3.4	&	--	&	--	&	--	&	14653	&	14584	&	14531	\\
3.6	&	--	&	--	&	--	&	14713	&	14646	&	14593	\\
3.8	&	--	&	--	&	--	&	14772	&	14706	&	14655	\\
4.0	&	--	&	--	&	--	&	14830	&	14765	&	14714	\\
4.2	&	--	&	--	&	--	&	--	&	14822	&	14772	\\
4.4	&	--	&	--	&	--	&	--	&	14878	&	14829	\\
4.6	&	--	&	--	&	--	&	--	&	14933	&	14885	\\
4.8	&	--	&	--	&	--	&	--	&	14987	&	14939	\\
5.0	&	--	&	--	&	--	&	--	&	15039	&	14993	\\
5.2	&	--	&	--	&	--	&	--	&	15091	&	15045	\\
5.4	&	--	&	--	&	--	&	--	&	15141	&	15096	\\
5.6	&	--	&	--	&	--	&	--	&	--	&	15147	\\
5.8	&	--	&	--	&	--	&	--	&	--	&	15196	\\
6.0	&	--	&	--	&	--	&	--	&	--	&	15244	\\
6.2	&	--	&	--	&	--	&	--	&	--	&	15292	\\
6.4	&	--	&	--	&	--	&	--	&	--	&	15339	\\
6.6	&	--	&	--	&	--	&	--	&	--	&	15385	\\
6.8	&	--	&	--	&	--	&	--	&	--	&	15430	\\
 \end{tabular} 
 \normalsize 
 \label{tab:tabrhovalues} 
 \end{ruledtabular}
 \end{table}

\section{Discussion}

\begin{figure}[ht]
  \includegraphics[width=\linewidth]{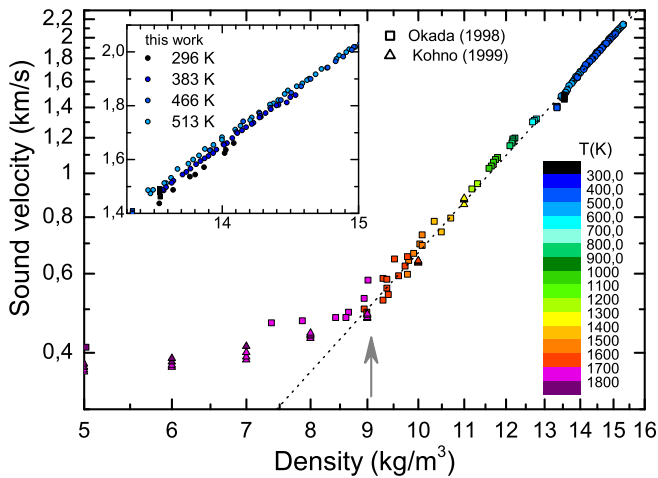}
  \caption{\label{fig:GraphKohno} Adiabatic sound velocity as a function of density at different temperatures in a log-log scale and linear scale (inset). 
  Sound velocity data from this work (circles) are plot together experimental data from Kohno~\cite{Kohno1999} (triangles up) and Okada~\cite{Okada1998} (squares). Grey arrow indicates the density at which metal - non metal transition occurs. The dashed line indicates a power law discussed in the text.}
\end{figure}

In figure~\ref{fig:GraphKohno} we report present and previous~\cite{Kohno1999} measurements of the adiabatic sound velocity as a function of density for different temperatures. The wide density range here reported spans from the low density non-metallic state to the high-density metallic one. As previously observed, in the metallic state the density is the most relevant parameter for determining the sound velocity, while pressure and temperature dependence of $v$ at constant densities is small. The effect of temperature on the density dependence of the sound velocity is shown in the insert of figure~\ref{fig:GraphKohno}. Conversely, on the nonmetallic side, $v$ varies slowly with density and it exhibits appreciable pressure (or temperature) dependence. This indicates that in the liquid metal the sound velocity can be described as a functional of density, and the different thermodynamic conditions produce only second order effects.

Near 9 g/cm$^3$, where the M-NM transition occurs, a clear inflection is observed due to the loose of the metallic character and the consequent modification of the interatomic interaction~\cite{Bomont2006}.
As can be observed, in the metallic state the sound velocity decreases rapidly with density with a decreasing rate of $\ln(v)/\ln(\rho)$ of the order of $2.72\pm0.02$\footnote{This value is lower than the value previously found by Okada \textit{et~al}~\cite{Okada1998} (this value was around 4-5).} (dashed line in figure~\ref{fig:GraphKohno}), which is also followed by our high pressure data.

In a crystal the propagation velocity of acoustic phonons can be directly linked to the derivatives of the pair potential. Due to the lack of translational invariance, it is not possible to formally write the same relation for the liquid. However empirical relations can be established which link the acoustic sound velocity to the effective interatomic interaction~\cite{March2005} in liquid metals.  

The scaling law here observed for the sound velocity as a function of density in the metallic state is thus indicative that pressures up to 7 GPa produce no significant change in the electronic density of the system and thus in the effective interatomic potential.

In conclusion, we performed accurate measurements of sound velocity in liquid mercury up to 7~GPa and temperatures up to 240\char23C using an original experimental method, the picosecond acoustics surface imaging in DAC.
We show that the thickness of the sample in a DAC can be accurately determined \textit{in-situ} by this technique as a function of pressure and temperature.
Using present velocities data, the density of the fluid is derived together with the pressure dependence of diffferent thermodynamic quantities as the bulk modulus or the heat capacity.

From a more general point of view, our study demonstrates that picosecond acoustics in DAC is a powerful technique to quantitatively extract sound velocity, density or thermodynamical quantities in liquid metals under extreme conditions. These state-of-the-art experiments will certainly be useful in several applied problems and many other fields such as geophysics.

\bibliography{base}

\end{document}